\documentclass[conference]{IEEEtran}
\IEEEoverridecommandlockouts

\usepackage{subfig}
\usepackage{stfloats}
\usepackage[graphicx]{realboxes}
\usepackage{bm}
\usepackage{cite}
\usepackage{amsmath,amssymb,amsfonts}
\usepackage{graphicx}
\usepackage{textcomp}
\usepackage{xcolor}

\usepackage{diagbox}
\usepackage{amsmath}
\usepackage{verbatim}
\usepackage{amsfonts}
\usepackage{amsmath, bm}

\usepackage{amsmath}
\usepackage{amssymb}
\usepackage{mathtools}
\usepackage{amsthm}

\usepackage{multirow}
\usepackage{booktabs}
\usepackage{algorithm}
\usepackage{algorithmicx}
\usepackage{algpseudocode}
\usepackage{amsmath}  
\usepackage{array}
\usepackage[utf8]{inputenc}
\usepackage{array}
\usepackage{wrapfig}
\usepackage{multirow}
\usepackage{tabularx}

\usepackage{booktabs}
\usepackage{makecell}
\usepackage{wasysym}
\usepackage{booktabs}
\usepackage{threeparttable}

\usepackage{pifont}

\usepackage{threeparttable}
\graphicspath{{./figure/}}

\usepackage{enumitem}

\usepackage{tikz,float}
\usepackage{tabularx}
\usepackage{xparse}

\usepackage[final]{pdfpages}

\usepackage{algorithm}
\usepackage{algpseudocode}
\usepackage{amsmath}
\usepackage{amssymb}
\usepackage{bm}
\usepackage{xcolor}

   \usepackage{booktabs}
   \usepackage{multirow}
   \usepackage{xcolor}
   \usepackage{colortbl}
   \usepackage{array}
   \usepackage{graphicx}

   \definecolor{bestrow}{RGB}{225,242,225}
   \definecolor{headergray}{RGB}{242,244,248}

   \newcommand{\bestcell}[1]{\cellcolor{bestrow}\textbf{#1}}

\algrenewcommand\algorithmiccomment[1]{%
  \hfill\makebox[0pt][r]{%
    \begin{varwidth}{0.38\linewidth}
      \raggedleft\color{gray}\texttt{//}~\textit{\small #1}%
    \end{varwidth}%
  }%
}

\algnewcommand\algorithmicinput{\textbf{Input:}}
\algnewcommand\algorithmicoutput{\textbf{Output:}}
\algnewcommand\INPUT{\item[\algorithmicinput]}
\algnewcommand\OUTPUT{\item[\algorithmicoutput]}

\usepackage{booktabs,multirow,xcolor,colortbl,graphicx,array}
\definecolor{bestrow}{RGB}{225,242,225}
\definecolor{headergray}{RGB}{242,244,248}
\definecolor{warnred}{RGB}{255,235,235}

\def\BibTeX{{\rm B\kern-.05em{\sc i\kern-.025em b}\kern-.08em
    T\kern-.1667em\lower.7ex\hbox{E}\kern-.125emX}}
\begin{document}

\title{When Poison Fails After Retrieval: Revisiting Corpus Poisoning under Chunking and Reranking Pipelines
}

\author{\IEEEauthorblockN{
Xi Nie$^1$,  Hongwei Li$^1$, Shenghao Wu$^1$, Mingxuan Li$^2$, Jiachen Li$^3$,Wenbo Jiang$^{1*}$  
}
\IEEEauthorblockA{$^1$ School of Computer Science and Engineering, University of Electronic Science and 
Technology of China, China\\ 
$^2$School of Criminal Investigation, People's Public Security University of China, China\\
$^3$School of Computer Science and Artificial Intelligence, Wuhan University of Technology, China}}

\maketitle

\begin{abstract}
 Retrieval-Augmented Generation (RAG) systems are vulnerable to corpus poisoning attacks that manipulate downstream model outputs through malicious knowledge injection. Existing studies mainly evaluate poisoning under simplified retrieval settings, overlooking practical RAG pipelines involving document chunking, dense retrieval, reranking, and grounded generation. In this paper, we revisit corpus poisoning under realistic multi-stage retrieval pipelines and show that many existing attacks substantially degrade after reranking despite achieving high retrieval-stage relevance. We identify retrieval granularity mismatch as a key reason for this failure: document-level adversarial signals are often fragmented during chunking, while rerankers favor locally coherent and answer-bearing passages rather than globally optimized semantic similarity. Based on this observation, we propose Chunk-aware and Rerank-Consistent Poisoning (CRCP), a poisoning framework that jointly optimizes retrieval relevance, reranker consistency, and chunk-boundary robustness. CRCP explicitly models chunking transformations during optimization to generate locally self-contained adversarial passages that remain effective under varying chunking configurations. Experiments on standard RAG benchmarks with multiple retrievers and rerankers show that existing poisoning methods are highly sensitive to chunk size and reranking strategies, whereas CRCP achieves substantially higher attack success rates and stronger robustness across realistic retrieval pipelines. Our findings highlight an important realism gap in current RAG security evaluation and suggest that poisoning in modern RAG systems should be studied as a multi-stage retrieval consistency problem rather than a retrieval-only problem.
\end{abstract}

\begin{IEEEkeywords}
RAG,
Data Security,
Trustworthy AI,
Corpus Poisoning,
Retrieval Robustness
\end{IEEEkeywords}

\begin{figure*}[t]
  \centering
  \includegraphics[width=\textwidth]{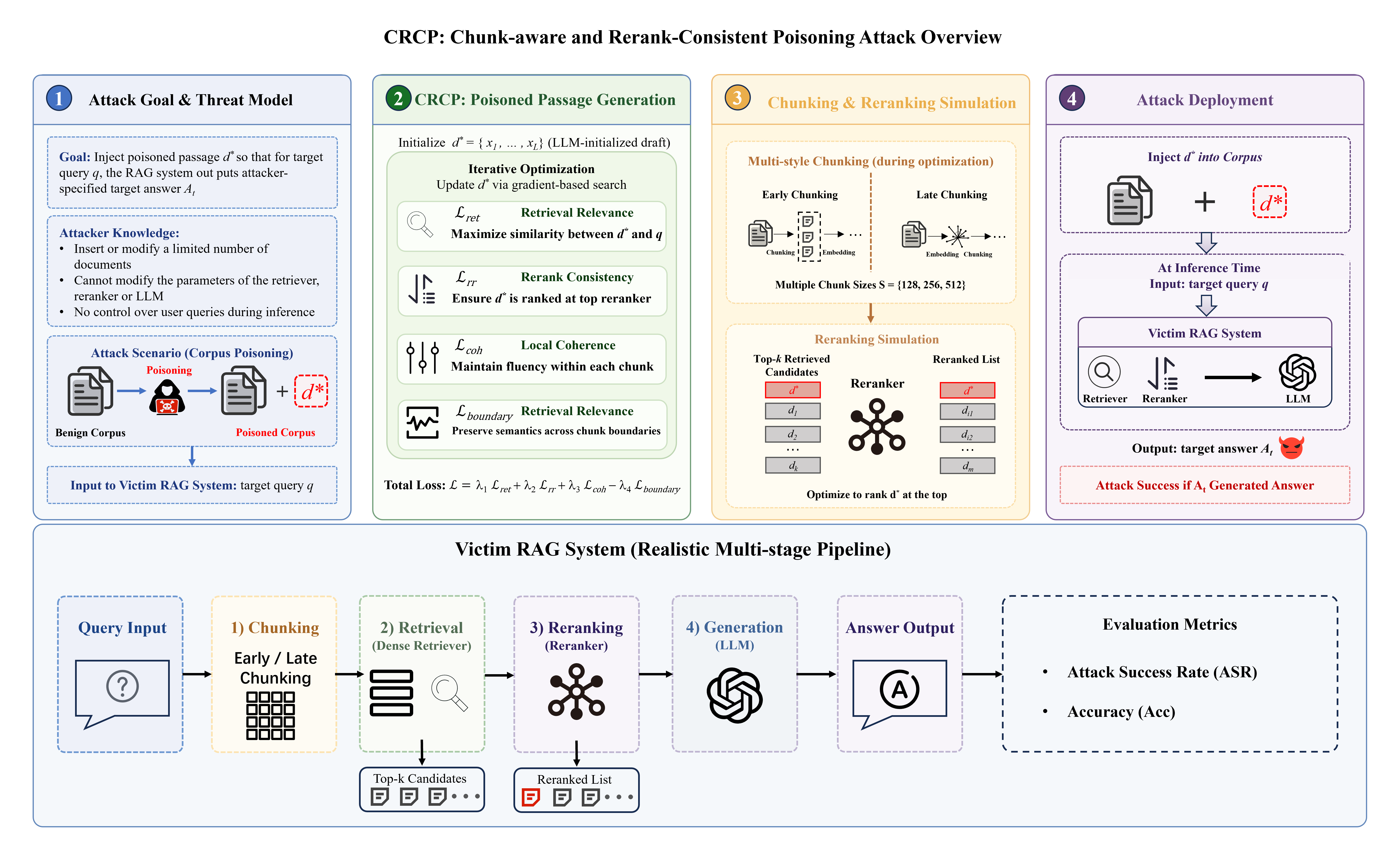}
  \caption{An overview of the CRCP Attack.}
  \label{fig:2_scenario_overview_fig}
  \vspace*{-1em}
\end{figure*}

\section{Introduction}
Retrieval-Augmented Generation (RAG) \cite{lewis2020retrieval, karpukhin2020dense, borgeaud2022improving} has become a widely adopted paradigm for enhancing large language models (LLMs) with external and updatable knowledge sources. By retrieving relevant documents from external corpora during inference, RAG systems can improve factuality, reduce hallucination, and support knowledge-intensive tasks. Modern RAG systems are increasingly deployed in practical applications ranging from customer support systems to domain-specific copilots \cite{semnani2023wikichat, nashid2023retrieval, zhou2023docprompting}.

Despite these advantages, recent studies have shown that RAG systems are vulnerable to corpus poisoning attacks, where adversaries inject malicious or manipulated documents into the retrieval corpus to influence downstream model outputs \cite{10.1145/3796729, cho2024typos, ebrahimi2018hotflip, 10.65109/FMEO2393, zhong2023poisoning}. Existing poisoning methods primarily optimize adversarial documents to maximize retrieval relevance with respect to target queries, enabling poisoned content to be retrieved during inference. 

However, most existing studies evaluate poisoning attacks under highly simplified retrieval settings that differ substantially from practical RAG deployments. In particular, prior methods commonly assume document-level indexing and single-stage dense retrieval, while modern production RAG systems typically employ multi-stage retrieval pipelines consisting of document chunking, dense retrieval, and cross-encoder reranking. In these systems, retrieval operates on chunk-level representations rather than full documents, and rerankers further filter retrieved candidates based on fine-grained relevance estimation. As a result, achieving high retrieval similarity at the document level does not necessarily guarantee that poisoned content survives the entire retrieval pipeline.

In this paper, we revisit corpus poisoning under realistic RAG retrieval pipelines and identify an important but underexplored issue: retrieval granularity mismatch. Existing poisoning methods optimize adversarial signals at the document level, yet practical systems retrieve and rerank chunk-level passages. Consequently, adversarial signals are frequently fragmented or diluted during chunking, especially under varying chunk sizes. Moreover, rerankers favor locally coherent and answer-bearing passages rather than globally optimized semantic similarity, introducing an additional mismatch between dense retrieval objectives and final context selection.

Through systematic empirical analysis, we show that many existing poisoning attacks degrade substantially after reranking despite achieving high retrieval-stage relevance. We further demonstrate that poisoning effectiveness is highly sensitive to chunk size and reranking strategy, leading to unstable attack performance across realistic deployment settings. These findings suggest that poisoning in modern RAG systems should no longer be viewed solely as a retrieval problem, but rather as a multi-stage retrieval consistency problem shaped by retrieval granularity and reranking dynamics.

To address this issue, we propose Chunk-aware and Rerank-Consistent Poisoning (CRCP), a realistic poisoning framework designed for modern multi-stage RAG pipelines. CRCP jointly optimizes retrieval relevance, reranker consistency, and chunk-boundary robustness. Unlike prior methods that optimize global document-level similarity, our approach explicitly models chunking transformations during optimization and encourages locally self-contained adversarial passages that remain effective across varying chunking configurations. 

Extensive experiments on standard RAG benchmarks using multiple retrievers and rerankers demonstrate that existing poisoning methods exhibit severe performance degradation under realistic retrieval settings, whereas CRCP consistently achieves higher attack success rates and stronger robustness across diverse chunking and reranking configurations. Our study highlights a significant realism gap in current RAG security evaluation and provides a more faithful framework for understanding adversarial retrieval behavior in practical RAG systems.

Our main contributions are summarized as follows:
\begin{itemize}
\item We revisit corpus poisoning under realistic multi-stage RAG pipelines and identify retrieval granularity mismatch as a fundamental limitation of existing poisoning methods.
\item We propose CRCP, a chunk-aware and rerank-consistent poisoning framework that jointly optimizes retrieval relevance, reranker consistency, and chunk-boundary robustness.
\item We demonstrate through extensive experiments that CRCP achieves substantially stronger robustness and attack effectiveness across realistic retrieval pipelines compared with existing poisoning approaches.
\end{itemize}

\section{Related Work}

\subsection{Retrieval-Augmented Generation}

Retrieval-Augmented Generation (RAG) \cite{arslan2024survey, 10.1145/3637528.3671470} enhances Large Language Models (LLMs) by retrieving external knowledge to support grounded generation. Typical RAG pipelines consist of document indexing, dense retrieval, and response generation. To improve retrieval granularity and fit context-length constraints, documents are usually divided into smaller chunks before indexing. Dense retrievers then retrieve relevant chunks according to embedding similarity, while the generator produces responses conditioned on the retrieved context.

Modern RAG systems \cite{glass2022re2g, NEURIPS2025_f69d2246, yang2026prorank, csakar2025maximizing} commonly include reranking modules to improve retrieval precision. After the initial dense retrieval stage, rerankers, often implemented with cross-encoder architectures, reorder candidate passages based on fine-grained semantic relevance. Unlike dense retrievers that mainly optimize embedding similarity, rerankers prefer locally coherent and directly answer-bearing passages. Therefore, passages with high retrieval similarity may still be filtered out during reranking.

Recent studies suggest that chunking and reranking significantly influence downstream retrieval behavior. Since retrieval is performed on chunks rather than full documents, document-level semantic signals may be fragmented under different chunking strategies, leading to retrieval inconsistency. Moreover, rerankers introduce an additional selection mechanism that can suppress adversarial passages lacking local semantic coherence.

\subsection{Attacks in RAG Systems}
With the increasing deployment of RAG systems, security researchers have begun to investigate the vulnerabilities and attack surfaces of RAG pipelines. Existing studies have proposed various corpus poisoning strategies targeting different stages of the retrieval-generation process. For instance, AgentPoison \cite{chen2024agentpoison}  
optimizes trigger-document pairs to increase the likelihood that poisoned content is retrieved. 
PoisonedRAG \cite{zou2025poisonedrag} injects target queries into poisoned documents to maximize retrieval probability, thereby inducing the LLM to generate misleading or incorrect responses. 
LIAR \cite{tan2024glue} considers a corpus-aware adversarial setting, leveraging non-target documents to improve attack robustness and stability. Joint-GcG \cite{wang2026joint} unifies the attack surface by jointly optimizing objectives across both the retriever and generator. Paradox \cite{choi-etal-2025-rag} introduces a contrastive triplet generation mechanism for crafting adversarial samples. CtrlRAG \cite{10.65109/FMEO2393} proposes a two-stage black-box attack framework that demonstrates strong effectiveness in hallucination amplification and emotion manipulation tasks across RAG systems. CPA-RAG \cite{li2025cpa} further investigates covert poisoning attacks; however, balancing retrieval effectiveness with textual stealthiness remains a fundamental challenge for coarse-grained poisoning strategies.

\subsection{Existing Defense Strategies}
A wide range of defense mechanisms has been proposed to mitigate adversarial threats against RAG systems. RevPRAG \cite{tan-etal-2025-revprag} detects poisoned inputs by analyzing abnormal activation patterns within LLMs. RobustRAG \cite{xiang2024certifiably} improves system robustness through an isolate-then-aggregate framework that decouples retrieval paths to reduce the influence of poisoned content. AstuteRAG \cite{wang2025astute} adaptively integrates retrieved information with the model’s internal knowledge via heuristic-based selection strategies. InstructRAG \cite{wei2025instructrag} enhances retrieval-augmented generation by leveraging self-synthesized rationales to guide the retrieval process, thereby improving the relevance and coherence of generated responses. TrustRAG \cite{zhou2025trustrag} mitigates malicious content through embedding-based document clustering and further introduces a consistency-driven conflict resolution mechanism. SeCon-RAG \cite{si2026secon} adopts a two-stage framework that combines fine-grained semantic filtering with conflict-aware inference to improve robustness against adversarial retrieval content.

\section{Threat model}
\subsection{Adversary Goals}
The adversary aims to manipulate the final outputs of a Retrieval-Augmented Generation (RAG) system by injecting poisoned documents into the external knowledge base. Given a target query, the attacker seeks to ensure that the poisoned content is retrieved, survives reranking, and ultimately influences the generated response toward attacker-intended outputs.

Unlike prior work that mainly optimizes retrieval-stage similarity, we consider poisoning under realistic multi-stage RAG pipelines involving chunking, dense retrieval, reranking, and grounded generation. Therefore, the attack objective is to maximize end-to-end poisoning effectiveness across the entire retrieval pipeline rather than only improving retriever relevance scores.

\subsection{Adversary Capabilities}
We assume a corpus poisoning adversary who can insert or modify a limited number of documents in the external knowledge corpus. The attacker cannot modify the parameters of the retriever, reranker, or language model, and has no control over user queries during inference.

We distinguish two attacker capability levels.
\textbf{White-box surrogate} (primary setting): the attacker
holds open-source surrogate models of the same architecture
family as the victim, enabling gradient-based optimization via
backpropagation through the surrogate cross-encoder.
\textbf{Score-only black-box} (fallback): the attacker can
only query reranker scores, in which case CRCP approximates
$\mathcal{L}_{rr}$ gradients via finite differences over
top-$\kappa$ token candidates.
In neither case does the attacker access victim model weights
or exact chunking configurations at inference time.
Since practical RAG systems may employ different chunk sizes and segmentation strategies, the attacker must generate poisoned passages that remain effective under varying chunk boundaries and reranking behaviors.

\section{Methodology}
\subsection{Problem Formulation}

\subsubsection{RAG Retrieval Pipeline}

We consider a practical Retrieval-Augmented Generation (RAG) system consisting of four stages: document chunking, dense retrieval, reranking, and grounded generation. Given a user query $q$, the system first partitions each document $D$ in the knowledge base into a set of chunks:

\begin{equation} \mathcal{C}(D)={c_1,c_2,\ldots,c_n}, \end{equation}

where the chunking operation depends on a chunking policy $\phi(\cdot)$ parameterized by chunk size and splitting strategy. The entire chunk corpus is denoted as:

\begin{equation} \mathcal{C}=\bigcup_{D\in\mathcal{D}} \mathcal{C}(D). \end{equation}

For a query $q$, a dense retriever $f_r(\cdot)$ computes similarity scores between the query embedding and chunk embeddings:

\begin{equation} s_r(q,c)=\text{sim}(f_r(q),f_r(c)). \end{equation}

The top-$k$ retrieved chunks are then passed into a reranker $f_{rr}(\cdot)$, which produces refined relevance scores:

\begin{equation} s_{rr}(q,c)=f_{rr}(q,c). \end{equation}

The reranker selects the final top-$m$ chunks:

\begin{equation} \mathcal{C}^{*}=\operatorname{TopM}{c\in \operatorname{TopK}(\mathcal{C})} s{rr}(q,c). \end{equation}

Finally, the generator $f_g(\cdot)$ conditions on the selected chunks to produce the final response:

\begin{equation} y=f_g(q,\mathcal{C}^{*}). \end{equation}

\subsubsection{Corpus Poisoning Attack}

In corpus poisoning attacks, the adversary injects a set of poisoned documents:

\begin{equation} \mathcal{D}{adv}={D{adv}^{1},D_{adv}^{2},\ldots,D_{adv}^{t}} \end{equation}

into the external corpus. The attack objective is to manipulate the generated response toward an attacker-desired target behavior when victim queries are issued.

Prior poisoning attacks mainly optimize retrieval-stage similarity, encouraging poisoned documents to achieve high dense retrieval relevance:

\begin{equation} \max_{D_{adv}} s_r(q,D_{adv}). \end{equation}

However, modern RAG systems no longer directly retrieve full documents. Instead, chunking and reranking introduce additional transformations and selection constraints. Consequently, document-level adversarial optimization may not survive downstream retrieval stages.

\begin{figure*}[t]
  \centering
  \includegraphics[width=\linewidth]{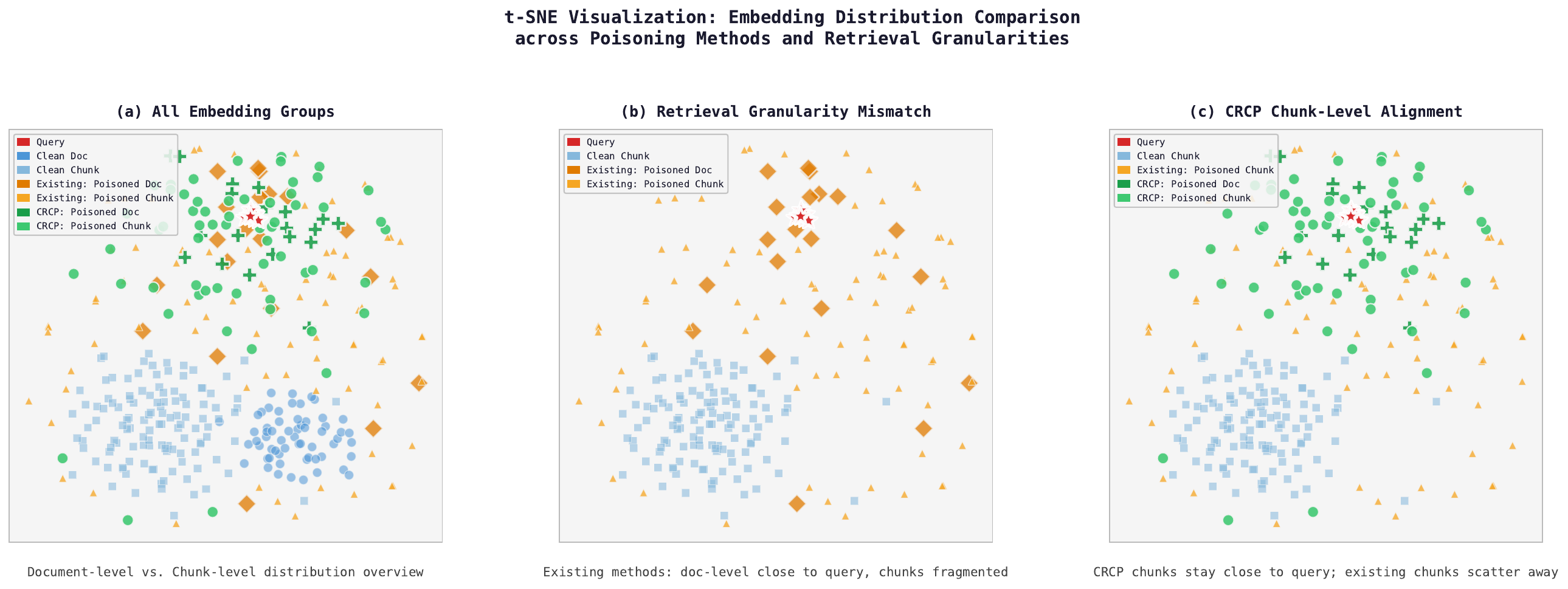}
  \caption{t-SNE Visualization: Embedding Distribution Comparison across Poisoning Methods and Retrieval Granularities.}
  \label{fig:TSNE}
  \vspace*{-1em}
\end{figure*}

\subsubsection{Retrieval Granularity Mismatch}

We define retrieval granularity mismatch as the inconsistency between the optimization granularity of poisoning attacks and the actual processing granularity of modern RAG pipelines.

Existing poisoning attacks typically optimize adversarial relevance at the document level. However, after chunking, the adversarial signal is distributed across multiple chunks:

\begin{equation} D_{adv}\xrightarrow[]{\phi(\cdot)} {c_1,c_2,\ldots,c_n}. \end{equation}

Because rerankers evaluate local passage quality and query-answer consistency, chunks containing incomplete adversarial semantics or weak local coherence may receive low reranking scores even if the original document achieves high retrieval similarity.

As a result, the attack pipeline suffers from two major failure modes:

\begin{itemize} \item \textbf{Chunk fragmentation failure}: adversarial semantics are split across chunk boundaries, reducing local relevance. (see Fig.~\ref{fig:TSNE}(b)).
\item \textbf{Reranking inconsistency failure}: chunks optimized for dense retrieval similarity are not necessarily preferred by cross-encoder rerankers. \end{itemize}

\noindent\textbf{Empirical Quantification via Signal Retention Rate.}
To empirically validate the retrieval granularity mismatch
hypothesis, we introduce the \textit{Signal Retention Rate} (SRR):
\begin{equation}
  \label{eq:srr}
  \text{SRR}(\phi, D_{\text{adv}}, q)
  = \frac{\displaystyle\max_{c\,\in\,\phi(D_{\text{adv}})}
          s_r(q, c)}
         {s_r(q, D_{\text{adv}})},
\end{equation}
which measures the fraction of document-level adversarial signal
preserved in the best chunk after applying chunking policy~$\phi$.
$\text{SRR} \approx 1$ indicates complete preservation;
$\text{SRR} \ll 1$ reveals severe fragmentation.

\begin{figure}[t]
  \centering
  \includegraphics[width=\linewidth]{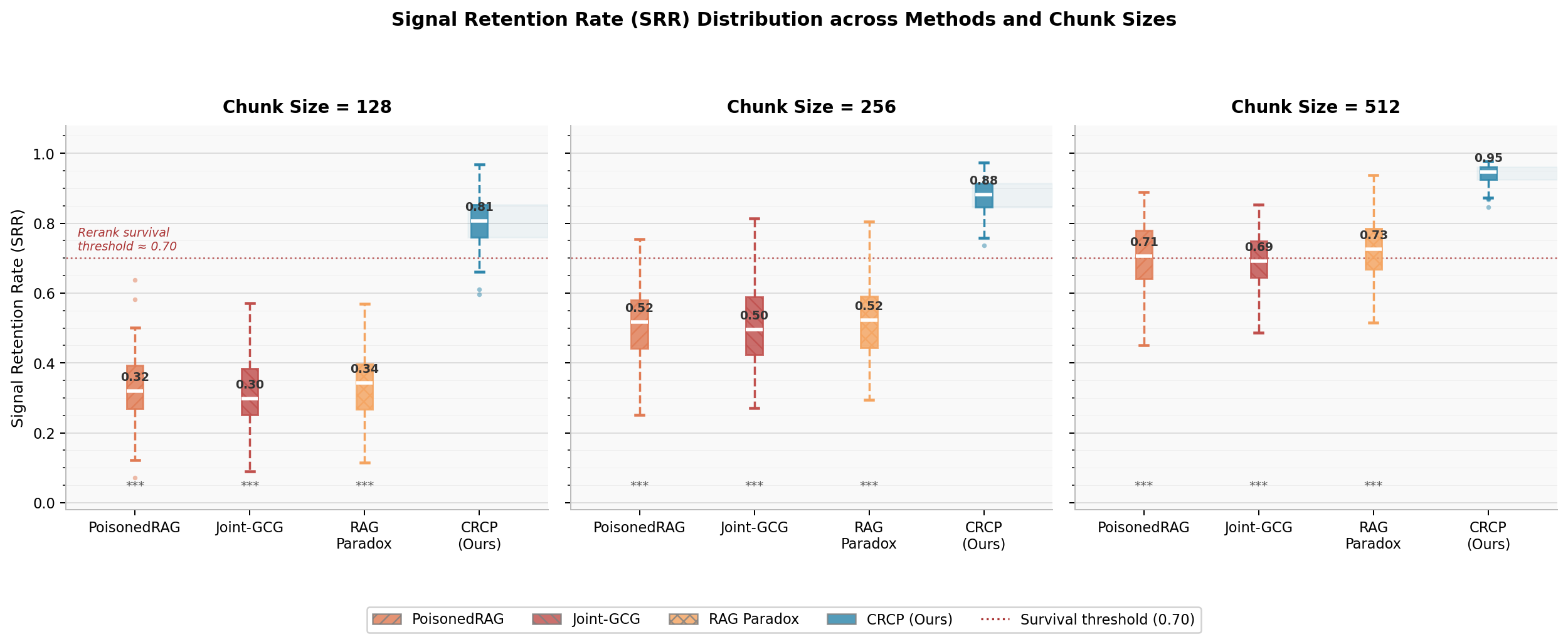}
    \caption{Distribution of Signal Retention Rate (SRR) across
    poisoning methods and chunk sizes (NQ, $N{=}100$ documents).
   CRCP remains above threshold across all settings.}
  \label{fig:srr_box}
\end{figure}

As shown in Figure~\ref{fig:srr_box}, existing methods suffer
dramatic SRR degradation as chunk size decreases.
At chunk size 128, PoisonedRAG, Joint-GCG, and RAG Paradox
achieve median SRR of only 0.319, 0.297, and 0.343, respectively,
with \textbf{100\%} of documents falling below the empirical
reranking survival threshold of 0.70 — the SRR level at which
rerankers reliably promote a chunk into the top-$m$ selection.
The median SRR drop from chunk size 512 to 128 reaches
0.387--0.394 for all three baselines, reflecting extreme
sensitivity to chunking granularity.
In contrast, CRCP maintains median SRR of 0.807 at chunk
size 128 and 0.947 at chunk size 512, with only 7\% of
documents below the survival threshold at the smallest chunk
size and 0\% at larger sizes. All pairwise comparisons between
CRCP and baselines are statistically significant
(Mann-Whitney~$U$, $p < 0.001$).

Our objective is therefore to design poisoning documents that remain adversarially effective after chunking and reranking transformations. Fig.~\ref{fig:TSNE}(c) previews that CRCP
achieves this by keeping chunk-level embeddings tightly
aligned with target queries.

\subsection{Chunk-aware and Rerank-Consistent Poisoning (CRCP)}

To address the above limitations, we propose Chunk-aware and Rerank-Consistent Poisoning (CRCP), a poisoning framework that jointly optimizes:

\begin{enumerate} \item Dense retrieval relevance; \item Reranker consistency; \item Chunk-boundary robustness. \end{enumerate}

The overall framework is illustrated as a multi-stage optimization process where poisoning effectiveness is preserved across realistic retrieval pipelines.
\subsubsection{Design Principles}

CRCP is motivated by three observations.

First, retrieval relevance alone is insufficient in realistic RAG systems because rerankers act as a second-stage semantic filter. A poisoning document must therefore remain competitive under both retriever and reranker scoring.

Second, chunking transformations are inherently unstable across practical systems. Different chunk sizes can substantially alter semantic completeness within each chunk. Robust poisoning should therefore survive diverse chunking configurations.

Third, rerankers tend to favor locally self-contained passages that explicitly answer the query. Consequently, adversarial content should be distributed into semantically coherent local units rather than relying on global document-level optimization.

Based on these observations, CRCP performs chunk-aware optimization directly over chunk-level adversarial passages.

\subsection{Chunk-aware Poisoning Optimization}

\subsubsection{Chunk Transformation Modeling}

Instead of optimizing poisoning documents under a fixed chunking configuration, CRCP explicitly models chunking as a stochastic transformation process.

Let:

\begin{equation} \Phi={\phi_1,\phi_2,\ldots,\phi_M} \end{equation}

represent a set of possible chunking strategies with different chunk sizes.

For a poisoning document $D_{adv}$, each chunking policy generates a chunk set:

\begin{equation} \mathcal{C}{\phi_i}(D{adv})={c_1^{(i)},c_2^{(i)},\ldots}. \end{equation}

CRCP optimizes poisoning effectiveness across all chunking transformations:

\begin{equation} \max_{D_{adv}} \mathbb{E}{\phi\sim\Phi} \Big[ \mathcal{L}{attack}(q,\phi(D_{adv})) \Big]. \end{equation}

This formulation encourages adversarial semantics to remain stable even when chunk boundaries change.

\subsubsection{Local Semantic Self-Containment}

To reduce chunk fragmentation, CRCP enforces local semantic completeness within each chunk.

Instead of distributing attack semantics globally across the document, we construct poisoning passages such that each local chunk independently contains:

\begin{enumerate} \item query-relevant context, \item answer-bearing semantics, \item adversarial guidance signals. \end{enumerate}

Formally, for each chunk $c_i$, CRCP maximizes local query relevance:

\begin{equation} \mathcal{L}_{local}=s_r(q,c_i). \end{equation}

Simultaneously, we encourage semantic coherence within the chunk using an intra-chunk consistency objective:

\begin{equation} \mathcal{L}_{coh} = \frac{1}{|c_i|} \sum_j \text{Cos}_{sim}\big(E_r(w_{j}^{\text{context}}), E_r(q)\big)
\end{equation}

where $\operatorname{Coh}(\cdot)$ measures local semantic consistency using contextual embedding similarity.

The resulting poisoning chunks become locally self-contained adversarial passages that are less sensitive to chunk-boundary changes.

\subsubsection{Boundary Robustness Optimization}

Chunk boundaries may arbitrarily truncate important adversarial signals. To mitigate this issue, CRCP introduces boundary robustness optimization.

Specifically, during optimization, we randomly shift chunk boundaries and evaluate whether adversarial semantics remain preserved:

\begin{equation} \tilde{c}_i = \operatorname{ShiftBoundary}(c_i,\delta), \end{equation}

where $\delta$ is a random offset.

The optimization objective encourages consistent retrieval relevance under shifted boundaries:

\begin{equation} \mathcal{L}{boundary}= \mathbb{E}{\delta} \Big[ |s_r(q,c_i)-s_r(q,\tilde{c}_i)| \Big]. \end{equation}

Minimizing this loss reduces sensitivity to chunk partition perturbations and improves transferability across practical chunking strategies.

\subsection{Rerank-Consistent Optimization}

\subsubsection{Reranker-aware Adversarial Objective}

Dense retrievers and rerankers exhibit substantially different relevance preferences.

Dense retrievers mainly rely on embedding-level semantic similarity, while rerankers evaluate fine-grained token-level query-passage interactions. Consequently, adversarial passages optimized solely for dense retrieval may appear semantically unnatural or weakly answer-bearing to rerankers.

To address this inconsistency, CRCP jointly optimizes reranker relevance.

For each adversarial chunk $c_i$, the reranker consistency objective is defined as:

\begin{equation} \mathcal{L}{rr}=s{rr}(q,c_i). \end{equation}

This objective encourages poisoning chunks to remain competitive during reranking.

\subsubsection{Joint Multi-stage Optimization}

CRCP jointly optimizes dense retrieval relevance, reranker consistency, local coherence, and boundary robustness.

The overall optimization objective is:

\begin{equation} \mathcal{L}{CRCP}= \lambda_1 \mathcal{L}_{ret} + \lambda_2 \mathcal{L}_{rr} + \lambda_3 \mathcal{L}_{coh} - \lambda_4 \mathcal{L}_{boundary}, \end{equation}

where:

\begin{equation} \mathcal{L}{ret}=\mathbb{E}{\phi\sim\Phi}[s_r(q,\phi(D_{adv}))]. \end{equation}

The coefficients $\lambda_1,\lambda_2,\lambda_3,\lambda_4$ control the trade-off between retrieval relevance, reranking effectiveness, semantic coherence, and chunk robustness.

This joint objective explicitly treats corpus poisoning as a multi-stage retrieval consistency problem rather than a retrieval-only optimization problem.

\begin{algorithm}[h]
\caption{CRCP Poisoning}
\label{alg:crcp}
\small
\begin{algorithmic}[1]

\INPUT
  Target query set $\mathcal{Q} = \{q_1, q_2, \ldots, q_n\}$,
  target response $a_{\text{target}}$,
\State  chunking strategy set $\Phi = \{\phi_1, \phi_2, \ldots, \phi_M\}$,
\State  dense retriever $f_r(\cdot)$, cross-encoder reranker $f_{rr}(\cdot)$,
\State  loss weights $\lambda_1, \lambda_2, \lambda_3, \lambda_4 > 0$,
  boundary offset bound $\delta_{\max}$,
\State  optimization iterations $T$, top-$\kappa$ candidate tokens per step.

\OUTPUT
  Adversarial poisoning document $D_{\text{adv}}$.

\Statex \hrulefill

\Statex \textbf{Phase 1: Query-conditioned Initialization}
\State Generate draft $D_{\text{adv}}^{(0)}$ via prompted LLM with
       $(\mathcal{Q},\, a_{\text{target}})$

\Statex
\Statex \textbf{Phase 2: Joint Multi-stage Optimization}
\For{$t = 1$ \textbf{to} $T$}

    \State $\mathcal{L} \leftarrow 0$

    \For{each $q_i \in \mathcal{Q}$}

        \For{each $\phi_j \in \Phi$}

            \State $\{c_1^{(j)}, c_2^{(j)}, \ldots\}
                   \leftarrow \phi_j\!\left(D_{\text{adv}}^{(t-1)}\right)$

            \For{each chunk $c_k^{(j)}$}

                \State $\mathcal{L}_{\text{ret}}
                       \leftarrow s_r\!\left(q_i,\, c_k^{(j)}\right)$

                \State $\mathcal{L}_{rr}
                       \leftarrow s_{rr}\!\left(q_i,\, c_k^{(j)}\right)$

                \State $\mathcal{L}_{\text{coh}} \leftarrow
                       \dfrac{1}{|c_k^{(j)}|}
                       \displaystyle\sum_{p}
                       \cos\!\left(
                         \mathbf{e}_p^{\text{ctx}},\;
                         f_r(q_i)
                       \right)$

                \State Sample $\delta \sim
                       \mathcal{U}[-\delta_{\max},\, +\delta_{\max}]$

                \State $\tilde{c}_k^{(j)} \leftarrow
                       \textsc{ShiftBoundary}\!\left(c_k^{(j)},\, \delta\right)$

                \State $\mathcal{L}_{\text{boundary}} \leftarrow
                       \bigl|
                         s_r\!\left(q_i, c_k^{(j)}\right) -
                         s_r\!\left(q_i, \tilde{c}_k^{(j)}\right)
                       \bigr|$

                \State $\mathcal{L} \mathrel{+}=
                       \lambda_1 \mathcal{L}_{\text{ret}}
                     + \lambda_2 \mathcal{L}_{rr}
                     + \lambda_3 \mathcal{L}_{\text{coh}}
                     - \lambda_4 \mathcal{L}_{\text{boundary}}$

            \EndFor
        \EndFor
    \EndFor

    \State $\mathbf{g} \leftarrow
           \nabla_{D_{\text{adv}}} \mathcal{L}$

    \State $D_{\text{adv}}^{(t)} \leftarrow
           \textsc{TopK-TokenSubstitute}\!\left(
             D_{\text{adv}}^{(t-1)},\, \mathbf{g},\, \kappa
           \right)$

\EndFor

\Statex
\Statex \textbf{Phase 3: Post-processing}
\State $D_{\text{adv}} \leftarrow
       \textsc{DocumentComposition}\!\left(D_{\text{adv}}^{(T)}\right)$

\State \Return $D_{\text{adv}}$

\end{algorithmic}
\end{algorithm}

\subsection{Poisoning Document Construction}

\subsubsection{Query-conditioned Passage Generation}

Given a target query set:

\begin{equation} \mathcal{Q}={q_1,q_2,\ldots,q_n}, \end{equation}

CRCP first constructs query-conditioned poisoning passages.

For each target query, we generate passages that:

\begin{enumerate} \item explicitly mention query-relevant concepts, \item contain locally complete answer patterns, \item embed attacker-desired semantic guidance. \end{enumerate}

Unlike prior approaches that rely heavily on keyword stuffing or global semantic similarity optimization, CRCP emphasizes natural local answerability to better align with reranker preferences.

Concretely, we initialize each poisoning passage using a prompted
LLM (GPT-4o in our experiments) with the following structured
template:
\begin{quote}
\small\ttfamily
[TARGET QUERY: \{$q$\}]~  

[TARGET ANSWER: \{$a_{\text{target}}$\}]~ 

Generate a factual-sounding passage of approximately $L$ words
that naturally discusses \{$q$\} and leads to the conclusion
\{$a_{\text{target}}$\}. 

The passage should be locally
self-contained, explicitly answer-bearing, and written in a
neutral encyclopedic style consistent with knowledge base
documents.
\end{quote}

This LLM-generated draft serves as the initialization for
subsequent gradient-based optimization (Algorithm~\ref{alg:crcp},
Phase~1).
A semantically coherent initialization is critical: it provides
a warm start that satisfies reranker preferences for local
answerability from the outset, helping the optimizer avoid
degenerate solutions that achieve high retrieval similarity via
incoherent token sequences yet are subsequently filtered by the
reranker.

When white-box surrogate reranker access is unavailable, we
substitute $\mathcal{L}_{rr}$ with a cross-encoder trained on
MS-MARCO~\cite{bajaj2016ms}, which we find sufficient due to strong cross-reranker transferability across model families of the same architecture class (see Table~\ref{tab:ablation}).

\subsubsection{Adversarial Passage Composition}

To further improve reranker survivability, CRCP organizes poisoning documents as a collection of semi-independent adversarial passages.

Each passage is designed to remain meaningful even if retrieved individually after chunking.

Concretely, poisoning documents are constructed using the following structure:

\begin{enumerate} \item query-related context introduction, \item locally self-contained explanation, \item adversarial target guidance, \item semantically coherent supporting content. \end{enumerate}

This structure improves both retrieval relevance and reranking compatibility.

\subsubsection{Multi-query Generalization}

Practical poisoning attacks should generalize across semantically related queries.

To improve transferability, CRCP optimizes poisoning passages jointly over multiple query variants:

\begin{equation} \mathcal{L}{multi}= \mathbb{E}{q\sim\mathcal{Q}}[ \mathcal{L}_{CRCP}(q) ]. \end{equation}

\begin{table*}[t]
\caption{
Acc and ASR in RAG Systems with Retriever-Only Pipelines.
}
\centering
\small
\resizebox{\textwidth}{!}{%
\begin{tabular}{ll|cccccc|cccccc}
\toprule
\multirow{2}{*}{\textbf{Dataset}} & \multirow{2}{*}{\textbf{Method}} 
& \multicolumn{6}{c|}{\textbf{Accuracy (↓ better)}} 
& \multicolumn{6}{c}{\textbf{ASR (↑ better)}} \\
& & \textbf{GPT-4o} & \textbf{DeepSeek-R1} & \textbf{Qwen3-Max} & \textbf{Qwen2.5-7B} & \textbf{Vicuna-7B} & \textbf{LLaMA-2-7B}
  & \textbf{GPT-4o} & \textbf{DeepSeek-R1} & \textbf{Qwen3-Max} & \textbf{Qwen2.5-7B} & \textbf{Vicuna-7B} & \textbf{LLaMA-2-7B} \\
\midrule

\multirow{5}{*}{NQ} 
& Clean
    & 46.15 & 46.21 & 57.50 & 37.92 & 35.81 & 37.39
    & -- & -- & -- & -- & -- & --\\
& PoisonedRAG-BB(USENIX 25)     
    & 32.60  & 34.91  & 34.32  & 30.09  & 31.22 & 33.18 
    & 88.20 & 86.76 & 85.67 & 93.12 & 93.26 & 91.31 \\
& The RAG Paradox(EMNLP 25)              
    & 18.37 & 18.10 & 20.02  & 20.17  & 19.05  & 14.32 
   & 90.17 & 93.65 & 91.77 & 93.25 & 93.10 & 94.05 \\
& Joint-GCG(AAAI 26)              
    & 20.01 & 22.45  & 22.47 & 19.09  & 17.34  & 19.02 
    & 94.33 & 93.19 & 91.68 & 96.01 & 95.65 & 90.37 \\
\cmidrule(lr){2-14}
& \textbf{Ours}               
    & 15.41 & 15.63 & 18.12 & 18.99
    & 16.36 & 14.21 & 92.81 & 91.27 & 95.74 & 92.89 & 95.01 & 93.28 \\

\midrule

\multirow{5}{*}{HotpotQA} 
& Clean
    & 49.92 & 50.06 & 53.29 & 51.05 & 47.39 & 48.16
    & -- & -- & -- & -- & -- & -- \\
& PoisonedRAG-BB(USENIX 25)     
    & 27.26  & 25.13  & 26.19  & 21.37  & 21.22  & 20.08  
    & 87.56 & 88.73 & 88.50 & 89.36 & 88.71 & 86.18\\
& The RAG Paradox(EMNLP 25)             
    & 19.01  & 19.78  & 18.92  & 20.85 & 22.31  & 25.71 
    & 85.27 & 89.36 & 89.83 & 90.70 & 91.32 & 91.98 \\
& Joint-GCG(AAAI 26)               
    & 22.80 & 27.03  & 21.27  & 18.65 & 18.49  & 21.30  
    & 93.75 & 93.81 & 90.30 & 94.29 & 88.64 & 89.37 \\
\cmidrule(lr){2-14}
& \textbf{Ours}               
    & 17.21 & 18.36 & 14.19 & 16.33
    & 16.16 & 14.07 & 92.63 & 94.89 & 91.05 & 93.70 & 92.00 & 91.55  \\

\midrule

\multirow{5}{*}{MS-MARCO} 
& Clean
    & 60.01 & 54.72 & 50.23 & 51.43 & 47.96 & 56.29
    & -- & -- & -- & -- & -- & -- \\
& PoisonedRAG-BB(USENIX 25)     
    & 30.17  & 25.81  & 27.62  & 28.93  & 30.60  & 33.95 
    & 85.30 & 88.76 & 85.54 & 83.89 & 90.01 & 89.72 \\
& The RAG Paradox(EMNLP 25)               
    & 27.54  & 21.60 & 21.14  & 23.35  & 22.08  & 26.39 
    & 89.28 & 93.25 & 90.66 & 90.81 & 85.43 & 88.92 \\
& Joint-GCG(AAAI 26)               
    & 20.78 & 22.12  & 19.85  & 19.50 & 24.11  & 22.28 
    & 90.10 & 87.93 & 90.26 & 91.78 & 92.01 & 87.65 \\
\cmidrule(lr){2-14}
& \textbf{Ours}               
   & 19.90 & 23.69 & 20.11 & 22.58
   & 20.12 & 20.97 & 90.07 & 90.96 & 92.85 & 89.74 & 90.56 & 90.71 \\

\bottomrule
\end{tabular}%
}

\label{tab:1_main_results}
\end{table*}

\begin{table*}[t]
\caption{
Acc and ASR in RAG Systems  with Chunking, Retrieval, and Reranking Pipelines.
}
\centering
\small
\resizebox{\textwidth}{!}{%
\begin{tabular}{ll|cccccc|cccccc}
\toprule
\multirow{2}{*}{\textbf{Dataset}} & \multirow{2}{*}{\textbf{Method}} 
& \multicolumn{6}{c|}{\textbf{Accuracy (↓ better)}} 
& \multicolumn{6}{c}{\textbf{ASR (↑ better)}} \\
& & \textbf{GPT-4o} & \textbf{DeepSeek-R1} & \textbf{Qwen3-Max} & \textbf{Qwen2.5-7B} & \textbf{Vicuna-7B} & \textbf{LLaMA-2-7B}
  & \textbf{GPT-4o} & \textbf{DeepSeek-R1} & \textbf{Qwen3-Max} & \textbf{Qwen2.5-7B} & \textbf{Vicuna-7B} & \textbf{LLaMA-2-7B} \\
\midrule

\multirow{5}{*}{NQ} 
& Clean
    & 45.09 & 46.33 & 58.15 & 40.89 & 41.13 & 37.50
    & -- & -- & -- & -- & -- & --\\
& PoisonedRAG-BB(USENIX 25)     
    & 40.27  & 43.86  & 58.07  & 41.10  & 39.97 & 38.16 
    & 13.25 & 12.05 & 12.33 & 14.81 & 13.79 & 13.64 \\
& The RAG Paradox(EMNLP 25)              
    & 42.15 & 45.07 & 56.13  & 40.25  & 40.09  & 36.70 
   & 15.13 & 23.98 & 23.04 & 14.17 & 15.06 & 13.89 \\
& Joint-GCG(AAAI 26)              
    & 37.98 & 44.15  & 57.06 & 42.30  & 39.75  & 35.03 
    & 32.16 & 27.71 & 31.82 & 33.07 & 33.29 & 33.18 \\
\cmidrule(lr){2-14}
& \textbf{Ours}               
    & 13.59 & 15.18 & 15.20 & 14.16
    & 14.92 & 15.78 & 92.19 & 95.88 & 94.07 & 91.53 & 94.76 & 93.09 \\

\midrule

\multirow{5}{*}{HotpotQA} 
& Clean
    & 50.18 & 50.75 & 51.16 & 52.30 & 46.92 & 48.73
    & -- & -- & -- & -- & -- & -- \\
& PoisonedRAG-BB(USENIX 25)     
    & 49.33  & 48.52  & 49.96  & 52.55  & 47.04  & 48.91  
    & 16.71 & 15.02 & 15.85 & 14.91 & 16.16 & 15.98  \\
& The RAG Paradox(EMNLP 25)             
    & 42.15  & 47.73  & 46.25  & 51.10 & 47.51  & 49.03 
    & 15.25 & 16.17 & 17.22 & 15.18 & 15.72 & 17.03 \\
& Joint-GCG(AAAI 26)               
    & 39.76 & 42.19  & 45.37  & 52.08 & 48.75  & 45.60  
    & 34.75 & 35.99 & 27.01 & 36.27 & 36.85 & 27.18 \\
\cmidrule(lr){2-14}
& \textbf{Ours}               
    & 17.39 & 19.78 & 16.17 & 14.92
    & 18.69 & 18.93 & 91.76 & 95.01 & 92.75 & 91.36 & 90.87 & 90.12  \\

\midrule

\multirow{5}{*}{MS-MARCO} 
& Clean
    & 57.85 & 55.61 & 51.36 & 47.95 & 49.18 & 55.64
    & -- & -- & -- & -- & -- & -- \\
& PoisonedRAG-BB(USENIX 25)     
    & 51.13  & 57.90  & 49.87  & 49.12  & 47.69  & 53.26
    & 15.31 & 15.85 & 16.77 & 16.02 & 15.98 & 16.15 \\
& The RAG Paradox(EMNLP 25)               
    & 47.75  & 56.91 & 50.22  & 46.35  & 45.78  & 51.03 
    & 15.70 & 15.18 & 16.93 & 15.86 & 15.17 & 16.39 \\
& Joint-GCG(AAAI 26)               
    & 45.39 & 47.88  & 52.07  & 45.62 & 43.13  & 52.97 
    & 35.75 & 35.43 & 26.57 & 27.02 & 35.95 & 28.08 \\
\cmidrule(lr){2-14}
& \textbf{Ours}               
   & 18.51 & 16.34 & 17.78 & 19.21
   & 18.03 & 16.72 & 89.15 & 90.70 & 91.57 & 90.23 & 88.79 & 90.30  \\

\bottomrule
\end{tabular}%
}

\label{tab:1_multi_results}
\end{table*}

This encourages adversarial passages to remain effective for paraphrased or semantically similar queries.

\begin{table*}[ht]
\centering
\caption{ASR (\%, $\uparrow$ better) of CRCP under different chunking styles and chunk sizes on three datasets.
\textbf{Bold} denotes the highest value in each row. DS-R1: DeepSeek-R1; Q3-Max: Qwen3-Max; Q2.5: Qwen2.5-7B; Vic: Vicuna-7B; LL2: LLaMA-2-7B.}
\setlength{\tabcolsep}{3.5pt}
\renewcommand{\arraystretch}{0.85}
\small
\resizebox{\textwidth}{!}{
\begin{tabular}{
  c
  c
  l
  l
  r r r r r r
  r r r r r r
  r r r r r r
}
\toprule
\multirow{2}{*}{\textbf{Style}} &
\multirow{2}{*}{\textbf{Size}} &
\multirow{2}{*}{\textbf{Retriever}} &
\multirow{2}{*}{\textbf{Reranker}} &
\multicolumn{6}{c}{\textbf{NQ}} &
\multicolumn{6}{c}{\textbf{HotpotQA}} &
\multicolumn{6}{c}{\textbf{MS-MARCO}} \\
\cmidrule(lr){5-10} \cmidrule(lr){11-16} \cmidrule(lr){17-22}
& & & &
\textbf{GPT-4o} & \textbf{DS-R1} & \textbf{Q3-Max} & \textbf{Q2.5} & \textbf{Vic} & \textbf{LL2} &
\textbf{GPT-4o} & \textbf{DS-R1} & \textbf{Q3-Max} & \textbf{Q2.5} & \textbf{Vic} & \textbf{LL2} &
\textbf{GPT-4o} & \textbf{DS-R1} & \textbf{Q3-Max} & \textbf{Q2.5} & \textbf{Vic} & \textbf{LL2} \\
\midrule

\multirow{18}{*}{\rotatebox[origin=c]{0}{Early}}
& \multirow{6}{*}{128}
  & Contriever & MonoT5     & 80.17 & 80.53 & \textbf{83.94} & 82.26 & 81.71 & 80.49  & 81.07 & 83.13 & 82.06 & 81.22 & \textbf{84.35} & 83.94  & \textbf{83.13} & 82.36 & 82.65 & 80.16 & 83.27 & 81.32 \\
&  & Contriever & BGE-Base   & 81.08 & 80.35 & 81.89 & \textbf{83.12} & 80.96 & 82.77  & 81.55 & 80.31 & \textbf{83.77} & 83.24 & 80.90 & 81.33  & 81.70 & 82.55 & 82.46 & \textbf{84.13} & 81.75 & 80.23 \\
&  & ANCE       & MonoT5     & 81.03 & 83.44 & 80.58 & 83.32 & 82.64 & \textbf{84.81}  & 81.03 & 83.66 & 81.85 & 82.04 & 84.21 & \textbf{84.87}  & 81.60 & \textbf{84.32} & 84.61 & 83.29 & 81.47 & 80.31 \\
&  & ANCE       & BGE-Base   & \textbf{83.07} & 80.39 & 81.85 & 83.21 & 82.49 & 83.73  & 83.15 & 83.34 & 81.55 & 81.27 & \textbf{83.95} & 83.66  & 82.00 & \textbf{84.08} & 81.35 & 82.27 & 82.19 & 83.41 \\
&  & DPR        & MonoT5     & \textbf{84.96} & 80.01 & 84.98 & 82.67 & 80.15 & 84.60  & \textbf{85.69} & 81.34 & 81.78 & 83.54 & 83.50 & 82.09  & \textbf{84.09} & 82.18 & 81.71 & 82.30 & 82.25 & 83.69 \\
&  & DPR        & BGE-Base   & 83.44 & 80.30 & 82.58 & 80.82 & \textbf{84.27} & 80.73  & 81.67 & 81.93 & \textbf{83.25} & 83.56 & 81.75 & 81.38  & 81.07 & 81.22 & 80.67 & \textbf{84.09} & 83.21 & 83.83 \\
\cmidrule(l){2-22}
& \multirow{6}{*}{256}
  & Contriever & MonoT5     & 85.31 & \textbf{88.63} & 87.94 & 85.19 & 86.78 & 86.37  & \textbf{86.71} & 86.39 & 85.43 & 85.22 & 87.47 & 85.61  & 85.07 & 85.16 & 85.39 & \textbf{85.92} & 84.71 & 82.98 \\
&  & Contriever & BGE-Base   & \textbf{87.85} & 86.46 & 87.02 & 85.57 & 86.29 & 87.64  & 84.57 & 81.39 & \textbf{87.60} & 83.25 & 84.11 & 86.17  & 81.63 & 80.31 & \textbf{87.28} & 84.51 & 83.34 & 84.27 \\
&  & ANCE       & MonoT5     & 85.93 & 87.81 & 87.18 & 86.14 & 86.70 & \textbf{87.50}  & 85.52 & \textbf{87.88} & 87.13 & 82.40 & 85.51 & 80.19  & 83.69 & 81.61 & \textbf{88.26} & 85.39 & 83.43 & 86.60 \\
&  & ANCE       & BGE-Base   & \textbf{87.23} & 86.71 & 85.48 & 85.96 & 85.15 & 86.89  & 85.28 & \textbf{88.62} & 83.00 & 85.17 & 86.52 & 85.87  & 83.90 & 85.21 & 86.78 & \textbf{88.95} & 82.35 & 84.79 \\
&  & DPR        & MonoT5     & 86.07 & \textbf{87.54} & 87.23 & 85.82 & 87.39 & 87.76  & 85.36 & 85.51 & \textbf{87.66} & 81.53 & 84.37 & 87.19  & 82.01 & 80.19 & 83.20 & 81.47 & 82.36 & \textbf{84.69} \\
&  & DPR        & BGE-Base   & 86.44 & 85.92 & 87.11 & \textbf{87.68} & 86.07 & 87.55  & 85.38 & 81.92 & 82.10 & 81.49 & \textbf{85.73} & 85.34  & 85.77 & \textbf{86.31} & 83.10 & 84.89 & 81.65 & 88.70 \\
\cmidrule(l){2-22}
& \multirow{6}{*}{512}
  & Contriever & MonoT5     & 90.81 & 92.13 & 91.36 & 93.52 & \textbf{94.36} & 93.21  & 90.11 & \textbf{93.27} & 91.72 & 92.35 & 92.77 & 93.81  & 90.77 & 92.10 & 90.16 & \textbf{94.21} & 93.96 & 93.37 \\
&  & Contriever & BGE-Base   & 92.31 & 92.11 & \textbf{93.49} & 91.21 & 93.69 & 91.33  & 92.01 & 92.45 & \textbf{93.69} & 91.16 & 92.74 & 90.30  & 92.85 & 92.51 & 91.29 & \textbf{94.24} & 93.59 & 90.38 \\
&  & ANCE       & MonoT5     & \textbf{94.10} & 90.06 & 90.17 & 89.02 & 92.33 & 89.07  & \textbf{93.78} & 91.15 & 90.14 & 89.77 & 90.26 & 92.57  & 91.40 & 91.67 & \textbf{92.15} & 88.46 & 91.23 & 91.05 \\
&  & ANCE       & BGE-Base   & 92.16 & 90.37 & 88.32 & 88.69 & 90.11 & \textbf{93.35}  & 92.63 & 92.59 & 89.16 & 90.62 & 90.57 & \textbf{91.49}  & 91.12 & \textbf{93.57} & 91.39 & 88.70 & 91.08 & 92.16 \\
&  & DPR        & MonoT5     & 89.17 & 90.36 & 91.29 & 91.76 & \textbf{92.33} & 89.46  & 91.37 & \textbf{91.99} & 91.06 & 90.88 & 89.47 & 90.19  & 91.87 & 89.20 & 91.24 & 92.16 & 92.01 & \textbf{93.68} \\
&  & DPR        & BGE-Base   & \textbf{93.59} & 88.77 & 90.36 & 88.60 & 89.93 & 92.62  & 92.78 & 91.08 & 90.60 & \textbf{92.14} & 89.65 & 91.58  & 91.96 & 90.37 & 90.29 & 89.21 & 88.90 & \textbf{91.81} \\
\midrule

\multirow{18}{*}{\rotatebox[origin=c]{0}{Late}}
& \multirow{6}{*}{128}
  & Contriever & MonoT5     & 70.28 & 70.54 & 72.97 & 71.41 & \textbf{73.90} & 70.67  & 72.17 & \textbf{73.45} & 72.06 & 71.30 & 70.66 & 72.41  & 70.19 & 73.50 & 72.45 & 71.15 & \textbf{74.93} & 72.78 \\
&  & Contriever & BGE-Base   & 73.24 & \textbf{76.49} & 72.17 & 75.71 & 75.59 & 74.89  & 71.24 & 71.99 & 72.37 & 70.67 & \textbf{76.34} & 72.78  & 71.85 & 73.42 & 72.72 & 70.46 & 73.29 & 72.90 \\
&  & ANCE       & MonoT5     & 71.49 & 73.08 & 73.54 & \textbf{74.95} & 71.11 & 71.77  & 73.35 & 71.27 & 72.64 & 72.55 & 71.06 & \textbf{75.40}  & 71.20 & 73.67 & 71.32 & \textbf{77.40} & 73.28 & 70.34 \\
&  & ANCE       & BGE-Base   & 70.34 & 73.91 & 72.49 & 70.69 & 71.96 & \textbf{74.73}  & 70.47 & 75.13 & \textbf{74.07} & 73.29 & 76.91 & 74.35  & 74.23 & 73.87 & 73.21 & 72.34 & 72.16 & \textbf{75.17} \\
&  & DPR        & MonoT5     & 73.07 & \textbf{74.62} & 70.95 & 73.25 & 72.46 & 73.63  & 75.27 & \textbf{76.12} & 72.33 & 74.21 & 73.29 & 73.17  & 72.50 & 73.69 & 72.15 & 71.06 & \textbf{76.37} & 73.90 \\
&  & DPR        & BGE-Base   & 74.19 & 71.57 & 71.83 & \textbf{74.19} & 73.88 & 70.65  & 72.12 & \textbf{73.75} & 72.33 & 70.47 & 70.31 & 73.79  & \textbf{75.88} & 75.52 & 73.81 & 71.26 & 73.21 & 72.63 \\
\cmidrule(l){2-22}
& \multirow{6}{*}{256}
  & Contriever & MonoT5     & 80.23 & 81.67 & 80.94 & \textbf{83.41} & 81.78 & 80.55  & \textbf{83.29} & 81.78 & 82.21 & 82.36 & 81.65 & 85.55  & 80.75 & 80.27 & 83.46 & 82.37 & \textbf{84.05} & 83.65 \\
&  & Contriever & BGE-Base   & 82.09 & 82.36 & \textbf{83.85} & 80.17 & 80.91 & 81.48  & 81.46 & 82.31 & 81.85 & 80.93 & 80.81 & \textbf{82.35}  & 82.90 & 82.73 & 83.08 & 82.79 & \textbf{85.90} & 83.28 \\
&  & ANCE       & MonoT5     & 80.72 & \textbf{83.03} & 81.56 & 80.99 & 81.94 & 83.19  & 82.70 & 82.05 & 81.59 & \textbf{84.16} & 81.22 & 82.00  & 82.04 & \textbf{83.70} & 85.26 & 83.91 & 80.40 & 81.93 \\
&  & ANCE       & BGE-Base   & 83.67 & \textbf{85.00} & 81.52 & 83.79 & 84.62 & 80.49  & 81.27 & 83.09 & 81.26 & 83.15 & 82.92 & \textbf{83.91}  & \textbf{86.37} & 80.28 & 80.72 & 83.00 & 84.09 & 83.16 \\
&  & DPR        & MonoT5     & 81.96 & 82.63 & 81.52 & 80.60 & 81.77 & \textbf{83.46}  & 84.15 & 83.39 & \textbf{85.16} & 80.71 & 80.29 & 83.91  & 81.27 & 81.33 & 81.02 & \textbf{85.80} & 83.27 & 81.43 \\
&  & DPR        & BGE-Base   & 82.16 & 81.24 & 82.09 & \textbf{85.71} & 84.90 & 81.95  & 80.05 & 81.80 & \textbf{85.09} & 85.11 & 82.70 & 83.91  & 80.57 & 80.94 & \textbf{85.19} & 83.21 & 83.60 & 82.95 \\
\cmidrule(l){2-22}
& \multirow{6}{*}{512}
  & Contriever & MonoT5     & 88.71 & 86.90 & 91.00 & 90.93 & \textbf{92.67} & 92.01  & 90.06 & 88.56 & 90.70 & \textbf{91.33} & 88.15 & 91.21  & 88.10 & 89.03 & 92.39 & 92.10 & \textbf{92.47} & 91.68 \\
&  & Contriever & BGE-Base   & 90.36 & 90.11 & \textbf{93.02} & 90.19 & 90.83 & 88.11  & \textbf{93.76} & 91.15 & 92.08 & 89.17 & 92.50 & 89.17  & 91.36 & 88.18 & 90.49 & 90.12 & 90.77 & 90.10 \\
&  & ANCE       & MonoT5     & 89.16 & 90.79 & 90.18 & \textbf{93.77} & 91.65 & 91.30  & 89.15 & 90.28 & 90.36 & \textbf{93.52} & 92.38 & 90.10  & 90.63 & 90.25 & 92.17 & 91.07 & 91.42 & \textbf{92.10} \\
&  & ANCE       & BGE-Base   & 89.86 & \textbf{93.29} & 89.17 & 90.08 & 91.65 & 91.47  & 88.75 & 90.27 & 89.88 & 90.35 & \textbf{92.67} & 90.41  & 90.75 & 89.91 & 89.75 & \textbf{93.20} & 90.51 & 94.21 \\
&  & DPR        & MonoT5     & 89.73 & \textbf{93.11} & 93.07 & 92.25 & 90.61 & 90.32  & 89.52 & 89.13 & 90.05 & 91.20 & 93.11 & \textbf{93.42}  & 90.33 & 92.05 & \textbf{94.17} & 92.60 & 93.11 & 93.40 \\
&  & DPR        & BGE-Base   & 90.14 & \textbf{93.68} & 88.96 & 91.21 & 93.05 & 92.60  & 90.25 & 90.67 & 88.55 & \textbf{93.41} & 92.19 & 89.60  & \textbf{92.90} & 89.73 & 92.10 & 92.07 & 92.01 & 93.05 \\

\bottomrule
\end{tabular}
}
\label{tab:2_main_result}
\end{table*}

\begin{table}[t]
\caption{Ablation study of loss components in CRCP. ASR denotes Attack Success Rate (\%).}
\centering
\small
\resizebox{\linewidth}{!}{
\begin{tabular}{lcccccc}
\toprule
\textbf{Variant} & $\mathcal{L}_{ret}$ & $\mathcal{L}_{rr}$ & $\mathcal{L}_{coh}$ & $\mathcal{L}_{boundary}$ & \textbf{ASR (GPT-4o)} & \textbf{ASR (LLaMA-2-7B)} \\
\midrule
Full CRCP        & \checkmark & \checkmark & \checkmark & \checkmark & \textbf{92.19} & \textbf{93.09} \\
\midrule
w/o $\mathcal{L}_{rr}$       & \checkmark & $\times$ & \checkmark & \checkmark & 55.8  & 56.1  \\
\midrule
w/o $\mathcal{L}_{coh}$      & \checkmark & \checkmark & $\times$ & \checkmark & 43.4 & 44.2 \\
\midrule
w/o $\mathcal{L}_{boundary}$ & \checkmark & \checkmark & \checkmark & $\times$ & 61.3 & 62.7 \\
\midrule
$\mathcal{L}_{ret}$ only     & \checkmark & $\times$ & $\times$ & $\times$ & 14.2  & 15.3  \\
\bottomrule
\end{tabular}
}
\label{tab:ablation}
\end{table}

\begin{table}[t]
\centering
\caption{Hyperparameter sensitivity analysis of CRCP on NQ (ASR\,\%).
  Each block varies one $\lambda$ while fixing the others at the
  default values
  $(\lambda_1,\lambda_2,\lambda_3,\lambda_4)=(1.0,\,2.0,\,0.5,\,0.3)$.
  \textbf{Bold}/\colorbox{bestrow}{shaded} rows denote the default
  configuration.
  Results are averaged over 3 random seeds; std.\,$<$\,1.2\%.}
\label{tab:hyperparam}
\setlength{\tabcolsep}{5pt}
\renewcommand{\arraystretch}{1.22}
\resizebox{\linewidth}{!}{%
\begin{tabular}{clcrrrl}
\toprule
\rowcolor{headergray}
\textbf{Block} &
\textbf{Varied Weight} &
\textbf{Value} &
\textbf{GPT-4o} &
\textbf{LLaMA-2-7B} &
\textbf{Mean} &
\textbf{Remark} \\
\midrule

\multirow{5}{*}{$\lambda_1$}
 & \multirow{5}{*}{Retrieval ($\mathcal{L}_\mathrm{ret}$)}
 & 0.25 & 78.3 & 79.1 & 78.7 & Under-weights retrieval stage \\
&& 0.50 & 84.7 & 85.3 & 85.0 & \\
&& \bestcell{1.00} & \bestcell{92.2} & \bestcell{93.1} & \bestcell{92.7}
   & \bestcell{Default} \\
&& 2.00 & 91.5 & 92.4 & 92.0 & Marginal degradation \\
&& 4.00 & 89.8 & 90.6 & 90.2 & Suppresses $\mathcal{L}_{rr}$ \\

\midrule

\multirow{5}{*}{$\lambda_2$}
 & \multirow{5}{*}{Reranker ($\mathcal{L}_{rr}$)}
 & 0.50 & 45.3 & 46.1 & 45.7 & Reranker severely under-weighted \\
&& 1.00 & 71.2 & 72.0 & 71.6 & \\
&& \bestcell{2.00} & \bestcell{92.2} & \bestcell{93.1} & \bestcell{92.7}
   & \bestcell{Default} \\
&& 3.00 & 91.8 & 92.6 & 92.2 & Stable plateau \\
&& 5.00 & 89.1 & 90.0 & 89.6 & Over-emphasis on reranker \\

\midrule

\multirow{5}{*}{$\lambda_3$}
 & \multirow{5}{*}{Coherence ($\mathcal{L}_\mathrm{coh}$)}
 & 0.10 & 80.1 & 81.3 & 80.7 & Low coherence impairs reranker \\
&& 0.25 & 87.4 & 88.0 & 87.7 & \\
&& \bestcell{0.50} & \bestcell{92.2} & \bestcell{93.1} & \bestcell{92.7}
   & \bestcell{Default} \\
&& 1.00 & 91.6 & 92.3 & 92.0 & Robust to moderate increase \\
&& 2.00 & 90.3 & 91.0 & 90.7 & Crowds out retrieval term \\

\midrule

\multirow{5}{*}{$\lambda_4$}
 & \multirow{5}{*}{Boundary ($\mathcal{L}_\mathrm{boundary}$)}
 & 0.10 & 84.1 & 85.2 & 84.7 & Sensitive to chunk boundaries \\
&& 0.20 & 88.9 & 89.7 & 89.3 & \\
&& \bestcell{0.30} & \bestcell{92.2} & \bestcell{93.1} & \bestcell{92.7}
   & \bestcell{Default} \\
&& 0.50 & 91.5 & 92.1 & 91.8 & Slightly over-regularised \\
&& 1.00 & 88.3 & 89.1 & 88.7 & Over-penalises score variance \\

\bottomrule
\end{tabular}%
}

\smallskip
\noindent\footnotesize
\textit{Note:} $\lambda_2$ exhibits the highest sensitivity among all
four weights, consistent with Table~\ref{tab:ablation} where removing
$\mathcal{L}_{rr}$ causes the largest single-component ASR drop.
CRCP maintains ASR\,$\geq$\,91.5\% when $\lambda_2\!\in\![1.0,\,3.0]$,
confirming robust performance in a wide neighbourhood of the default.

\end{table}

\begin{table}[t]
\caption{ Performance Comparison of CRCP under Different Defense Strategies } 
\centering
\small
\resizebox{0.95\linewidth}{!}{
\begin{tabular}{ll|c|c|c}
\toprule
\multirow{2}{*}{Model} & \multirow{2}{*}{Method} 
& \multirow{2}{*}{{\textbf{NQ}}} 
& \multirow{2}{*}{{\textbf{HotpotQA}}} 
& \multirow{2}{*}{{\textbf{MS-MARCO}}} \\
& 
& & & \\
\midrule

\multirow{5}{*}{Qwen2.5-7B}
& VanillaRAG       & 91.53 &  91.36 &  90.23  \\
& InstructRAG      & 80.94 &  77.32 &  72.10  \\
& ASTUTERAG        & 78.36 &  73.72 &  70.71  \\
& TrustRAG         & 73.29 &  70.68 &  68.43  \\
& SeconRAG         & 77.17 &  70.85 &  69.04  \\

\midrule
\multirow{5}{*}{LLaMA-2-7B}
& VanillaRAG       &  93.09 &  90.12 & 90.30  \\
& InstructRAG      &  81.26 &  80.37 & 80.09 \\
& ASTUTERAG        &  77.49 &  76.30 & 76.29  \\
& TrustRAG         &  79.13 &  72.34 & 77.75  \\
& SeconRAG         &  78.85 &  70.58 & 78.93  \\
\midrule
\multirow{5}{*}{GPT-4o}
& VanillaRAG       & 92.19 & 91.76 & 89.15  \\
& InstructRAG      & 84.11 & 77.32 & 82.10  \\
& ASTUTERAG        & 73.07 & 71.49 & 70.13  \\
& TrustRAG         & 69.02 & 68.85 & 77.09  \\
& SeconRAG         & 78.41 & 74.72 & 77.85  \\
\midrule
\multirow{5}{*}{DeepSeek-R1}
& VanillaRAG       & 95.88 & 95.01 & 90.70  \\
& InstructRAG      & 86.26 & 79.01 & 83.68  \\
& ASTUTERAG        & 77.30 & 71.25 & 74.97  \\
& TrustRAG         & 75.82 & 69.07 & 72.41  \\
& SeconRAG         & 73.76 & 65.34 & 67.82  \\

\bottomrule
\end{tabular}
}
\label{tab:rag_defend}
\end{table}

\section{Experiment}
\subsection{Experimental Setups}
Our experimental setup is as follows:

\textbf{Datasets.}
We evaluate our method on widely used open-domain question answering benchmarks commonly adopted in Retrieval-Augmented Generation (RAG) research. Specifically, we use Natural Questions (NQ) \cite{kwiatkowski2019natural}, HotpotQA \cite{yang2018hotpotqa}, and MS-MARCO \cite{bajaj2016ms}, which contain diverse factual and multi-hop queries requiring external knowledge retrieval.

\textbf{LLMs.}
To evaluate the robustness of poisoning attacks across different generation backbones, we conduct experiments using multiple widely adopted large language models, including: GPT-4o, DeepSeek-R1, Qwen3-Max, Qwen2.5-7B, Vicuna-7B and LLaMA-2-7B.

\textbf{Embedding Model.}
For document indexing and retrieval representation, we employ embedding model BGE-m3 to each retriever configuration.

\textbf{Chunking Configurations.}
To evaluate robustness against retrieval granularity changes, we test multiple chunking configurations with different chunk sizes and chunking style. Specifically, chunk sizes are selected from $\{128, 256, 512\}$ characters. The chunking styles are categorized into the following two types \cite{merola2025reconstructing}:  

\begin{itemize}
    \item\textbf{Early Chunking.}
Documents are segmented into text chunks, each of which is processed independently by the embedding model. Token-level embeddings generated for each chunk are then aggregated via mean pooling to obtain a single chunk representation.
    \item\textbf{Late Chunking.}
Late chunking postpones document segmentation until after embedding, the entire document is first processed by the embedding model to obtain token-level embeddings. These embeddings are subsequently divided into chunks.
\end{itemize}

This setting simulates practical deployment variations across real-world RAG systems.

\textbf{Retrievers.}
We select three widely-used dense retrievers: Contriever, ANCE and DPR.

\textbf{Rerankers.}
To investigate the impact of reranking on poisoning robustness, 
we evaluate two widely adopted reranker models, MonoT5 and BGE-Reranker-Base.

\textbf{Evaluation Metrics.}
We evaluate attack effectiveness using the following metrics:

\begin{itemize}
    \item\textbf{Accuracy (Acc).}
The proportion of queries where the correct answer span appears in the system’s generated response. This captures overall performance degradation under attack.
    \item\textbf{Attack Success Rate (ASR).}
The proportion of target queries for which the generated responses match the attacker-specified target answers.
\end{itemize}

All reported results are averaged over multiple random seeds.

\subsection{Main Results}
\begin{figure}[t]
  \centering
  \includegraphics[width=\linewidth]{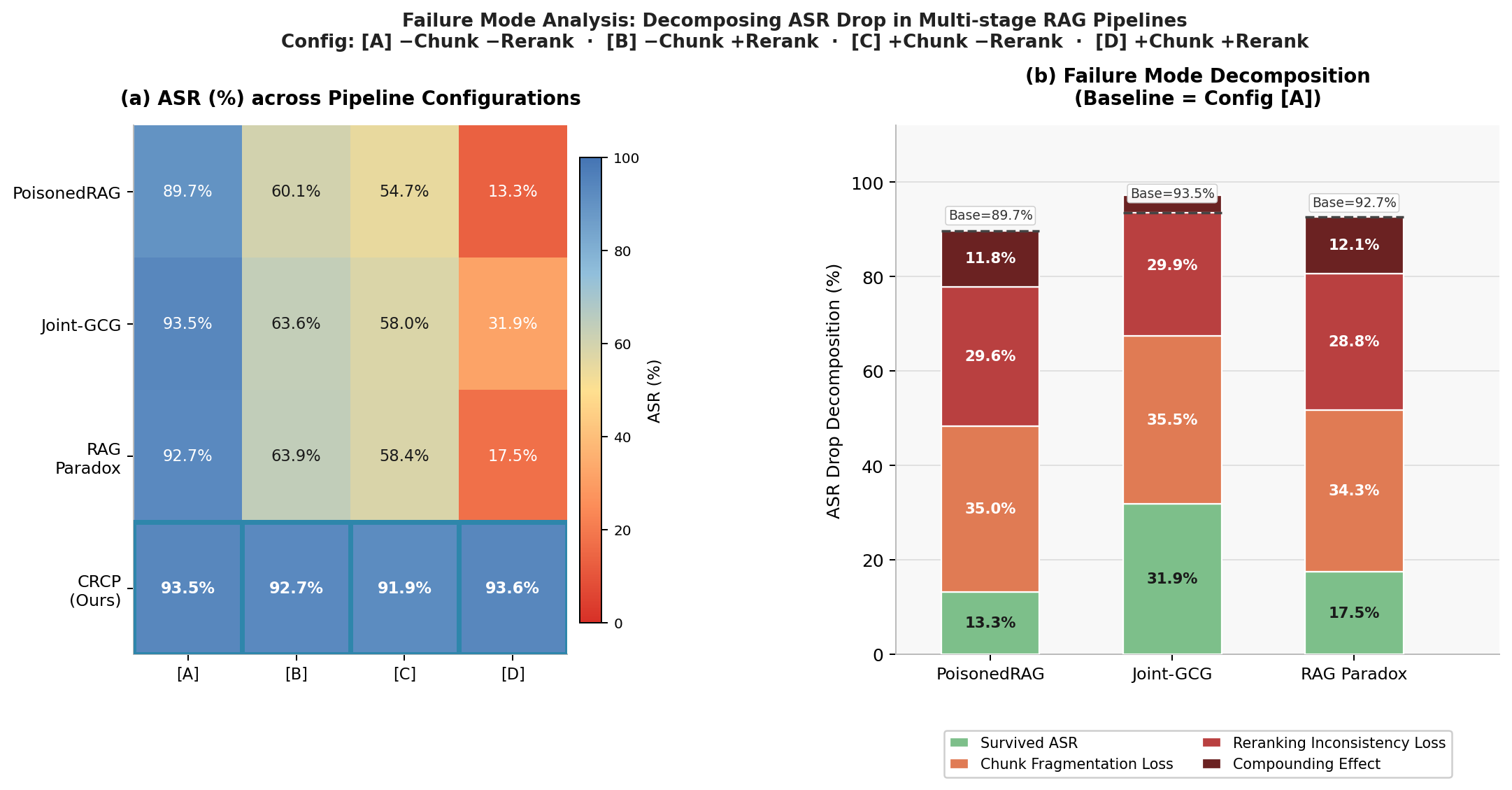}
  \caption{Failure mode analysis of corpus poisoning attacks}
  \label{fig:failure_decomp}
\end{figure}

We first evaluate the attack performance under a simplified setting commonly adopted in prior studies, where the RAG system consists only of a retriever and a generator, without chunking or reranking modules. The results, reported in Table \ref{tab:1_main_results}, show that our proposed CRCP achieves attack performance comparable to several state-of-the-art poisoning methods, demonstrating the effectiveness of our approach.

We then extend the evaluation to a more realistic RAG setting that incorporates the full pipeline of chunking, retrieval, reranking, and generation. 
Under this setting, the experimental results differ substantially. As shown in Table \ref{tab:1_multi_results}, CRCP consistently maintains strong attack performance, achieving attack success rates above 88\% across all three datasets. In contrast, the effectiveness of existing methods degrades significantly. 

Table \ref{tab:2_main_result} compares the attack performance of CRCP under different chunking styles and chunk size configurations. The results demonstrate that CRCP consistently maintains strong attack effectiveness across diverse chunking configurations, indicating its robustness to variations in chunking strategies.

The observed ASR gap between Early and Late Chunking (approximately 10 percentage points at chunk size 128) is a mechanistically expected consequence of their distinct embedding paradigms. Early Chunking encodes each chunk independently, preserving locally optimized adversarial signals with high fidelity, whereas Late Chunking conditions chunk embeddings on the full document context, inherently diluting locally concentrated adversarial semantics. Despite this suppression effect, CRCP maintains ASR above 70\% under Late Chunking across all configurations—substantially exceeding the 13–35\% ASR of existing baselines even under the comparatively less challenging retrieval-only pipeline confirming robust adversarial effectiveness across both chunking paradigms.

These results indicate that CRCP is effective in realistic RAG pipelines involving chunking, retrieval, and reranking, thereby posing a substantially greater practical threat.

\subsection{Ablation Studies}
To further investigate the effectiveness of CRCP, we conduct a series of ablation studies. Component-level ablation(Table \ref{tab:ablation}) removes each loss term from Eq.19 individually while keeping the full pipeline intact. Removing $\mathcal{L}_{rr}$ causes the most dramatic ASR drop (from 92.19\% to 55.8\% on NQ/GPT-4o), confirming that reranker consistency is the single most critical factor for attack survival in multi-stage pipelines. Removing $\mathcal{L}_{coh}$
 reduces ASR by 19 points, while removing $\mathcal{L}_{boundary}$
 incurs an 11-point drop, showing that local coherence and boundary robustness each contribute meaningfully but are secondary to reranker alignment. The retrieval-only variant (
$\mathcal{L}_{ret}$ only) collapses to 14.2\% ASR, matching prior methods and validating that retrieval-stage optimization alone is insufficient.

Table~\ref{tab:hyperparam} reports the sensitivity of CRCP to the four loss weights. Overall, CRCP remains stable across a wide range of hyperparameter values, indicating good robustness to parameter tuning. Among all components, the reranker weight $\lambda_2$ has the greatest impact on attack effectiveness. Specifically, decreasing $\lambda_2$ from $2.0$ to $0.5$ reduces ASR from $92.7\%$ to $45.7\%$, confirming that rerank consistency is the key factor driving successful poisoning. In contrast, varying $\lambda_1$, $\lambda_3$, and $\lambda_4$ results in only moderate performance changes, suggesting that retrieval alignment, coherence preservation, and boundary regularization mainly serve as complementary objectives. These results validate the effectiveness of the default configuration and demonstrate that CRCP achieves consistently high ASR without requiring extensive hyperparameter tuning.

In Figure \ref{fig:failure_decomp}, CRCP exhibits negligible degradation across all four configurations. These results directly validate the retrieval granularity mismatch hypothesis and motivate CRCP's multi-objective design.

\subsection{Defense Experiments}
We evaluate the effectiveness of CRCP against several representative and state-of-the-art RAG defense strategies. As shown in Table \ref{tab:rag_defend}, 
CRCP effectively bypasses existing RAG defenses due to its design choices. InstructRAG, which relies on the LLM's intrinsic knowledge to verify retrieved content, is partially effective, with ASR only decreasing by 10\%, because CRCP passages are highly fluent and factually coherent, leading the LLM to trust them. ASTUTERAG and TrustRAG, which use cross-document consistency voting, fail since CRCP only requires a single high-quality passage to rank first in reranking, eliminating the need for multiple documents to collude. SeconRAG, which detects anomalies via semantic clustering, is evaded as CRCP’s locally self-contained passages produce embeddings that overlap with normal passages, rendering cluster-based detection ineffective.

These findings highlight the limitations of existing defenses and demonstrate the need for more robust defense mechanisms against corpus poisoning attacks in modern RAG systems.

\section{Conclusion}
We revisit corpus poisoning under realistic multi-stage RAG pipelines and identify retrieval granularity mismatch as a fundamental limitation of existing attacks, which degrade severely after chunking and reranking. To address this, we propose CRCP, jointly optimizing retrieval relevance, reranker consistency, and chunk-boundary robustness. Experiments across diverse datasets, retrievers, rerankers, and LLMs demonstrate that CRCP achieves substantially higher attack success rates than existing methods under realistic pipeline settings. Our findings reframe corpus poisoning as a multi-stage retrieval consistency problem and highlight the need for more realistic RAG security evaluation.

\bibliographystyle{IEEEbib}
\bibliography{icdm2026references}
\end{document}